\documentclass[12pt,oneside,openany]{article}
\usepackage{authblk}
\usepackage{amsfonts}
\usepackage{amssymb}
\usepackage{amsthm}
\usepackage{amsmath}
\usepackage[]{algorithm2e}
\usepackage{algorithmic}
\usepackage{fullpage}
\usepackage{cite}

\def\binary{\lbrace 0,1 \rbrace}
\def\binaryn{\lbrace 0,1 \rbrace ^ n}
\def\binaryinfty{ \lbrace 0,1 \rbrace ^ \infty}
\def\binarystar{\lbrace 0,1 \rbrace ^ *}

\def\martingale{\binarystar \longrightarrow [0,\infty)}
\def\DNFwidth{\mathrm{DNF}_\mathrm{width}}

\def\NTIME[#1]{ \mathrm{NTIME($#1$)}}
\def\NTIME{\mathrm{NTIME}}

\def\DTIME{\mathrm{DTIME}}
\def\P{\mathrm{P}}
\def\NP{\mathrm{NP}}
\def\EXP{\mathrm{EXP}}
\def\E{\mathrm{E}}
\newcommand{\p}{\mathrm{p}}
\newcommand{\ptwo}{\mathrm{p}_2}

\newcommand{\myset}[2]{ \left\{ #1 \left| #2 \right.\right\} }

\newtheorem{theorem}{Theorem}[section]
\newtheorem{corollary}[theorem]{Corollary}
\newtheorem{lemma}[theorem]{Lemma}

\title{\bf Nondeterminisic Sublinear Time Has Measure 0 in P}

\author{John M. Hitchcock}
\author{Adewale Sekoni}
\affil{Department of Computer Science\\University of Wyoming}
\date{}

\begin{document}
\maketitle

\begin{abstract}
   The measure hypothesis is a quantitative strengthening of the $\P
   \neq \NP$ conjecture which asserts that $\NP$ is a nonnegligible
   subset of $\EXP$. Cai, Sivakumar, and Strauss (1997) showed that
   the analogue of this hypothesis in $\P$ is false. In particular,
   they showed that $\NTIME[n^{1/11}]$ has measure 0 in $\P$. We
   improve on their result to show that the class of all languages
   decidable in nondeterministic sublinear time has measure 0 in
   $\P$. Our result is based on DNF width and holds for all four major
   notions of measure on $\P$.
\end{abstract}
   
\section{Introduction}
   A central hypothesis of resource-bounded measure
   \cite{Lutz:AEHNC,Lutz:QSET} is that $\NP$ does not have measure 0
   in $\EXP$
\cite{Lutz:CvKL,Lutz:TPRBM}.  Cai, Sivakumar, and Strauss
\cite{cai1997constant} proved the surprising result that
$\NTIME[n^{1/11}]$ has measure 0 in $\P$.  This implies the analogue
of the measure hypothesis in $\P$ fails, because $\NTIME[\log n]$ has
nmeasure 0 in $\P$.

   We improve the result of Cai et al. by showing that the class of
   all languages that can be decided in nondeterministic time at most
   $$n\left(1-\frac{2\lg\lg n} {\lg n}\right)$$ has measure 0 in $\P$.
   In particular, the nondeterministic sublinear time
   class $$\NTIME[o(n)]$$ has measure 0 in $\P$.

Resource-bounded measure was initially defined for exponential-time
and larger classes \cite{lutz1992almost}. Defining measure within
subexponential- and polynomial-time complexity classes has been
challenging \cite{allender1994measure} and there are several notions
\cite{strauss1997measure,moser2008martingale} The result of Cai et
al. holds for a notion of measure on $\P$ we will refer to as
$\Gamma_d(\P)$-measure. Moser \cite{moser2008martingale} developed a
new notion of measure called $F$-measure.  It is the only notion of
measure that allows for defining resource-bounded dimension
\cite{lutz2003dimension} at $\P$.
It was unknown whether or not the result of Cai et al. also holds for
$F$-measure. Our result holds for $\Gamma(\P)$ measure (defined in
\cite{allender1994measure}) and therefore for $F$-measure and all the
notions of measure at P considered in
\cite{strauss1997measure,moser2008martingale}.
 
Our stronger result also has a much easier proof than the proof in
\cite{cai1997constant}.  Cai et al. use H\aa stad's switching lemma and
pseudorandom generators to show that the set of languages with nearly
exponential size circuits has $\Gamma_d(\P)$-measure 0
\cite{cai1997constant}. We use DNF width rather than the circuit size
to improve their result. It is well known that a random Boolean
function has DNF width close to $n$ (see \cite{CramaHammer01}). In
Section \ref{sec:DNFwidth}, we show that the class of languages with
sublinear DNF width has measure 0 in P. This is then applied in
Section \ref{sec:NTIME} to show that nondeterministic sublinear time
also has measure 0 in P.

\section{Preliminaries}\label{sec:prelim}

\subsection{Languages and Boolean functions}
   The set of all binary strings is $\lbrace 0,1 \rbrace ^*$.
   The length of a string $x\in\binarystar$ is denoted by $|x|$.
   The empty string is denoted by $\lambda$. For $n\in\mathbb{N}$, $\binaryn$ 
   is the set of strings of length $n$. $s_0=\lambda, s_1=0, s_2=1, s_3=00,
   ... $ is the standard lexicographic enumeration of
   $\binarystar$. A language $L$ is a subset of $\binarystar$. The set of 
   length $n$ strings of a language $L$ is $L^{=n} = L \cap \binaryn$. 
   Associated with every language $ L $ is its 
   characteristic sequence $\chi_L \in \lbrace 0,1 \rbrace ^ \infty $. It is  
   defined as \[\chi_L[i] = 1 \iff s_i \in L \ \textrm{for} \ i \in 
   \mathbb{N},\] where $\chi_L \lbrack i \rbrack $ is the $ i^{th} $ bit of $ 
   \chi_L $. We also index $\chi_L$ with strings i.e. for $i\in\mathbb{N}$,
   $\chi_L [s_i] = \chi_L [i]$. $\chi_L[i,j]$ denotes the $i^{th}$ through 
   $j^{th}$ bits of $\chi_L $, while $\chi_L[\textrm{length } n]$ 
   denotes $\chi_L[2^n - 1, 2^{n+1}-2]$, i.e. the substring of the 
   characteristic string of $L$ corresponding to the strings in $ L^{=n} $.
   
   A Boolean function is any $f:\binaryn \longrightarrow \binary$. A DNF
   (disjunctive normal form) formula of $f$ over the variables $x_1,x_2,\cdots,
   x_n$ is the logical OR of terms. A term is a logical AND of literals, where 
   a literal is either a variable $x_i$ or its logical negation
   $\overline{x}_i$. We require that no term contains a 
   variable and its negation \cite{o2014analysis}. Also the logical OR of the 
   empty term computes the constant $\mathbf{1}$ function while the the empty 
   DNF computes the constant $\mathbf{0}$ 
   function. A term's width is the number of literals in it. The size of a DNF 
   computing $f$ is the number of terms in it, while its width is the length of 
   its longest term. The DNF width of $f$ is the shortest width of any DNF 
   computing $f$. We note that the width of the constant $\mathbf{0}$ and $
   \mathbf{1}$ functions is 0. For any term $T$ we say
   that $T$ fixes a bit position $i$ if either $x_i$ or its negation appear in
   $T$. The bit positions that aren't fixed by $T$ are called free bit 
   positions. For example the term $x_1x_3\bar{x}_4:\binary^4 \longrightarrow 
   \binary$, fixes the first, third and fourth bit positions, while the second
   bit position is free. We say that $T$ covers a subset of $\binaryn$ if it
   evaluates to true on only the elements of the subset. The subset
   covered by $T$ is the set of all strings that agree with $T$ on all its
   fixed bit positions. A string $x \in \binaryn$ agrees with $T$ if, for any 
   fixed bit position $i$ of $T$, the $i$th bit of $x$ is $1$ if and only if 
   $x_i$ appears in $T$. We call the subset covered by $T$ a subcube of 
   dimension $n-k$, where $k$ is the number of literals in $T$. It is called a 
   subcube because it is a dimension $n-k$ hamming cube contained in the 
   dimension $n$ hamming cube.
   
   Associated with any Boolean function is its characteristic string
   $\chi_f \in \binary^{2^n}$ defined as
   \[f(w) = 1 \iff \chi_f[w] = 1 \textrm{ for } w \in \binaryn. \]
   For any language $L$ we view $L^{=n}$ as the Boolean function
   $\chi_{L^{=n}}$ defined as
   \[\chi_{L^{=n}}(w) = 1 \iff L[w] = 1 \textrm{ for all } w\in\binaryn.\]
   We can then define $\DNFwidth(L^{=n})$ to be the DNF width of $\chi_{L^{=n}}$.

\subsection{Resource-bounded Measure at $\P$}
   
Resource-bounded measure was introduced by Lutz
\cite{lutz1992almost}. He used martingales and a resource bound
$\Delta \supseteq \p$ to characterize classes of languages as either
``big" or ``small". Here $\p$ is the class of functions computable in
polynomial time.
Resource-bounded measure is a generalization of classical Lebesgue
measure.  For a given resource bound $\Delta \supseteq \p$ we get a
``nice" characterization of sets of languages as having measure 0,
measure 1 or being immeasurable with respect to $\Delta$.  Associated
with each resource bound $\Delta$ is a class $R(\Delta)$ that does not
have $\Delta$-measure 0. We can then use $\Delta$-measure to define a
measure on classes within $R(\Delta)$. For example, $\p$-measure yields
a measure on the exponential-time class $R(\p) = \E = \DTIME[2^{O(n)}]$. For the class
$\ptwo$ of quasipolynomial-time computable functions, $\ptwo$-measure
yields a measure on $R(\ptwo) = \EXP = \DTIME[2^{n^{O(1)}}]$.  See
\cite{ambos1997resource,Lutz:QSET} for a survey of
resource-bounded measure in $\Delta \supseteq \p$.

An apparently more difficult task is developing a notion of
resource-bounded measure on subexponential classes, in particular
developing a measure on $\P$ \cite{allender1994measure}.
There are at least four notions of measure defined on $\P$. Three of
these are due to Strauss \cite{strauss1997measure} and one is due to
Moser \cite{moser2008martingale}. None of them are quite as ``nice" as
measures on $R(\Delta) \supseteq \E$, each one of them having some
limitations. See
\cite{strauss1997measure,allender1995measure,moser2008martingale} for
a more detailed discussion of the limitations of these notions of
measure. In this paper we only consider one notion of measure on $\P$
we call $\Gamma(\P)$-measure.  $\Gamma(\P)$-measure was introduced in
\cite{allender1994measure}. We use $\Gamma(\P)$-measure for two
reasons. First, it is the simplest of the four notions of measure on
P. Second, the martingales considered in $\Gamma(\P)$-measure can be
easily shown to be martingales in the other notions of measure at
$\P$ \cite{strauss1997measure,moser2008martingale}.

\subsection{$\Gamma(\P)$-measure}
   A martingale is a function $d:\martingale$ that satisfies the the following 
   averaging condition:
   \[d(w) = \frac{d(w1)+d(w0)}{2}, \forall w \in \binarystar.\]
   Intuitively, the input $w \in \binarystar$ to the martingale $d$ is a 
   prefix of the characteristic sequence of a language. The martingale starts 
   with initial capital $d(\lambda)$. More generally, $d(w)$ is the 
   martingale's current capital after betting on the strings
   $s_0, s_1, \cdots, s_{|w|-1}$ in the standard ordering. The martingale $d$ 
   tries to predict the membership of string $s_{|w|}$ when given input $w$. If 
   $d$ chooses to bet on $s_{|w|}$ and is successful in predicting its
   membership, then its current capital increases, otherwise it decreases. The 
   martingale $d$ can also choose to not risk its current capital $d(w)$ by not 
   betting on $s_{|w|}$. The goal is to make $d$ grow without bound on some 
   subset of $\binaryinfty$. We say a martingale $d$ succeeds on a language $L$ 
   if \[\limsup_{n \rightarrow \infty}d(\chi_L[0,n-1]) = \infty.\]
   We say $d$ succeeds on a class $C \subseteq \binaryinfty$ if it succeeds on 
   every language in $C$. It is easy to see that the probability a martingale 
   $d$ succeeds on a randomly selected language is 0. A language $L$ is 
   randomly selected by adding each string to $L$ with probability 1/2. It can 
   be shown that any class $C \subseteq \binaryinfty$ has measure 0 under the 
   probability measure if and only if some martingale $d$ succeeds on $C$.
   If $d$ can be computed in some resource bound $\Delta$ then we say that $C$
   has $\Delta$-measure 0 if $d$ succeeds on $C$.
   
   A $\Gamma(\P)$-martingale is a martingale $d$ such that:
   \begin{itemize}
     \item  $d(w)$ can be computed by a Turing machine $M$ with oracle access 
        to $w$ and input $s_{|w|}$. We denote this computation as
        $M^w(s_{|w|})$.
     \item $M^w(s_{|w|})$ is computed in time polynomial in $\lg(|w|)$. In 
        other words, the computation is polynomial in the length of $s_{|w|}$.
     \item $d$ only bets on strings in a $\P$-printable set denoted $G_d$.
   \end{itemize}
  
   The input string $s_{|w|}$ to $M^w(s_{|w|})$ allows the Turing
   machine to compute the length of $w$ without reading all of $w$
   whose length is exponential in the length of $s_{|w|}$. A set $S
   \subseteq \binarystar$ is $\P$-printable \cite{AllRub88} if
   $S\cap\binaryn$ can be printed in time polynomial in $n$. A class $C
   \subseteq \binaryinfty$ has $\Gamma(\P)$-measure 0 zero if there is
   some $\Gamma(\P)$-martingale that succeeds on it
   \cite{strauss1997measure}.

\section{Measure and DNF Width}\label{sec:DNFwidth}

In this section we show that the class of languages with sublinear
DNF width has measure 0 in $\P$. Recall that for a language $L$,
$\DNFwidth(L^{=n})$ denotes the DNF width of the characteristic
string of $L$ at length $n$.

   \begin{theorem}
   \label{alphaClami}
      The class
      \[X = \myset{L \in \binaryinfty}{\ \DNFwidth(L^{=n}) \leq n \left(1-\frac{2\lg\lg n}
      {\lg n}\right)\mathrm{ i.o.}}\]
      has $\Gamma(\P)$-measure 0.
      \begin{proof}
         For clarity we omit floor and ceiling functions.
         \subsection*{The Martingale}
         Consider the following martingale $d$ that 
         starts with initial capital $4$. Let $L$ be the language $d$ is 
         betting on. $d$ splits its initial capital capital into portions 
         $C_{i,1}, C_{i,2}, i \in \mathbb{N}$, where $C_{i,1} = C_{i,2}
         = 1/n^2$. $C_{n,1}$ and $C_{n,2}$ are reserved for betting on strings 
         in $\binaryn.$ For each length $n$, $d$ only risks $C_{n,1}$ and
         $C_{n,2}$. Thus, $d$ never runs out of capital to bet on
         $\binaryn$ for all $n \in \mathbb{N}$. 
         
         Now we describe how $d$ bets on $\binaryn$ with $C_{n,1}$. $d$ 
         uses $C_{n,1}$ to bet that the first $n$ strings of $\binaryn$ 
         don't belong to $L$. If $d$ makes no mistake then the capital 
         $C_{n,1}$ grows from $1/n^2$ to $2^n/n^2.$ But once $d$ 
         makes a mistake it loses all of $C_{n,1}$, i.e. $C_{n,1}$ becomes
         $0.$
         
         Next we describe how $d$ bets on $\binaryn$ with $C_{n,2}$. The 
         martingale $d$ only bets with capital $C_{n,2}$ if it loses $C_{n,1}$, 
         i.e. the martingale $d$ makes a mistake on the first string of length 
         $n$ that belongs to $L$. Let us call this 
         string $w$. Let $w_1,w_2,\cdots,w_{n/\lg n}$ be a partition of $w$ 
         into $n/\lg n$ substrings, such that $w=w_1w_2\cdots w_{n / \lg n}$ 
         and each $w_i$ has length $\lg n$. %
         Furthermore $d$ splits
         $C_{n,2}$ into ${\lg n \choose 2\lg\lg n} \frac{n}{\lg n}$ 
         equal parts, i.e. $n^{1+o(1)}$ many parts. We 
         refer to each of them as $C_{n,2,i}, \textrm{ for } i \in [1,2,\cdots,
         {\lg n \choose2\lg\lg n}]$. Each one is reserved for betting according 
         to the prediction of some dimension $2\lg\lg n$ subcube that contains 
         $w.$ We only consider subcubes containing $w$ whose free bits lie 
         completely in one of the $w_i$'s. Let us call these subcubes the 
         boundary subcubes of $w$. It is easy to see that there are
         ${\lg n \choose 2\lg\lg n} \frac{n}{\lg n}$ boundary subcubes of $w$.
         
         Finally, to completely specify $d$, we describe how it bets with 
         each $C_{n,2,i}$ on any string $x \in \binaryn$ that comes after $w$, 
         the string $d$ lost all of $C_{n,1}$ on. $d$ bets as follows:
   
         \begin{algorithm}[H]
         \For{each boundary subcube $B_i$ of $w$}{
            $C_{n,2,i} \gets $ \textrm{capital reserved for betting on} $B_i$\;
            \If{$x \in B_i$}{
               verify that if $y < x$ and $y \in B_i$, then $y \in L$\;
               proceed to next $B_i$ if the verification fails\;
               bet all of $C_{n,2,i}$ on $x$ being in $L$\;
            }
         } 
         \caption{How $d$ bests on $x\in\binaryn$ that comes after $w$.}
         \end{algorithm}
         
         Intuitively, each $C_{n,2,i}$ is reserved for betting on a
         boundary subcube of $w$. The martingale predicts that each subcube
         is contained in $L^{=n}$. If the subcube $B_i$ which contains $w$ is 
         really contained in $L^{=n}$, then the capital reserved for betting on 
         this subcube grows from $C_{n,2,i}$ to $2^{2^{2\lg\lg n}-1}C_{n,2,i}$.
         This follows because we don't make any mistakes while betting on the 
         $2^{2\lg\lg n}-1$ strings in $B_i\setminus\{w\}$, and each of these 
         bets doubles $C_{n,2,i}$.
         \subsection*{The Martingale's Winnings on $X$}
         We now show that $d$ succeeds on any $L \in X$ by examining 
         its winnings on $L^{=n}$. In the first case, suppose the first $n$ 
         strings of $\binaryn$ are all not contained in $L$. In this case we bet 
         with $C_{n,1}$ and raise this capital from $1/n^2$ to $2^n/n^2$. In 
         the second case, suppose
         $\DNFwidth(L^{=n}) \leq n(1-\frac{2\lg\lg n}{\lg n})$ and one 
         of the first $n$ strings of $\binaryn$ is in $L$. Let us denote the 
         first such string by $w$. In this case $d$ will lose all of $C_{n,1}$
         and have to bet with $C_{n,2}$. Since
         $\DNFwidth(L^{=n}) \leq n(1-\frac{2\lg\lg n}{\lg n})$,
         $w$ must be contained in a subcube of dimension at least
         $(\frac{2\lg\lg n}{\lg n})n$. By a simple averaging argument it can 
         be seen that there must be at least one boundary subcube of $w$ that
         has dimension at least $2\lg\lg n$. Since $d$ must bet on such a 
         subcube, its capital reserved for this subcube rises from $C_{n,2,i}$ 
         to $2^{2^{2\lg\lg n}-1}C_{n,2,i}=\Theta(n^{\lg n})$. Since any
         $L\in X$ satisfies the above two cases infinitely often, $d$'s
         capital rises by $\Omega(n^{\lg n})$ infinitely often. Thus, $d$ 
         succeeds on $X$.
         
         \subsection*{The Martingale is a $\Gamma(\P)$-Martingale}
         Now we need to show $d$ is a $\Gamma(\P)$-martingale. It is easy 
         to see that $d$ is computable in time polynomial in $n$. Since for 
         each $x \in \binaryn$ we bet on, we iterate though $n^{1+o(1)}$ 
         sububes of dimension $2\lg\lg n$, and each subcube contains
         $O(\lg^2n)$ points. Also the set of strings that $d$ bets on in 
         $\binaryn$ is $\P$-printable since it only bets on the $n^{2+o(1)}$ 
         points in the boundary subcubes of the first $n$ strings of length
         $n$.
      \end{proof}
   \end{theorem}

   \section{Measure and Nondeterministic Time}\label{sec:NTIME}
   
   The following lemma is a generalization of an observation made in 
   \cite{cai1997constant}.
   
   \begin{lemma}
   \label{claimNTime}
      If $L$ can be decided by a nondeterministic Turing machine in time
      $f(n) \leq n$, then $L$ has DNF width at most
      $f(n)$.
      \begin{proof}
         We will show that for all %
$n$, $L^{=n}$ is
         covered by a DNF of width at most $f(n).$ If $L^{=n} = \emptyset$, 
         then it is covered by the empty DNF which has width 0. All that's 
         left is to show that $L^{=n}$ is covered by subcubes of dimension at 
         least $n-f(n)$ whenever $L^{=n} \neq \emptyset$. This is sufficient 
         because every subcube of dimension at least $n-f(n)$ is covered by a 
         width $f(n)$ term, so $L$ can be covered by a width $f(n)$ DNF. 
         Let $M$ be a nondeterministic Turing machine that decides 
         $L$ in time at most $f(n)$ and $x \in L^{=n}$.
         Thus, there is a nondeterministic computation of $M$ on input $x$ that 
         accepts. Since $M$ uses at most $f(n)$ time it can only examine at
         most $f(n)$ bits of $x$. So there are at least $n-f(n)$ bits of $x$ 
         that aren't examined by $M$ on some accepting computation of $M$ on 
         $x$. Therefore the set of all strings $y \in \binaryn$ that agree 
         with $x$ in all the bit positions examined by an accepting 
         computation must also be accepted by the same computation. This set
         of strings is precisely a subcube of dimension at least $n-f(n)$; 
         therefore, it is covered by a DNF term of width at most $f(n)$. Since 
         $x \in L^{=n}$ was arbitrary, it follow that $L^{=n}$ can be covered 
         by DNF term(s) of width at most $f(n)$; therefore, $L^{=n}$ has DNF 
         width at most $f(n)$. 
      \end{proof}
   \end{lemma}

   \begin{theorem}
      \label{NTtheorem}
      The class of all languages decidable in nondeterministic time at most 
      $n(1-\frac{2\lg\lg n}{\lg n})$ infinitely often has $\Gamma(\P)$-measure 
      0.
      \begin{proof}
         By lemma \ref{claimNTime}, any language decidable in nondeterministic 
         time at most $n(1-\frac{2\lg\lg n}{\lg n})$ has DNF width at most
         $n(1-\frac{2\lg\lg n}{\lg n})$ for all but finitely many $n$. 
         Therefore it follows by theorem \ref{alphaClami} that the set of all 
         such languages have $\Gamma(\P)$-measure 0.
      \end{proof}
   \end{theorem}

We now have the main result of the paper:

\begin{corollary} $\NTIME\left[n(1-\frac{2\lg\lg n}{\lg n})\right]$ has
  $\Gamma(\P)$-measure 0.
\end{corollary}

\begin{corollary} $\NTIME[o(n)]$ has $\Gamma(\P)$-measure 0.
\end{corollary}

Because $\Gamma(\P)$ measure 0 implies measure 0 in the other notions
of measure on $\P$ \cite{strauss1997measure,moser2008martingale},
Theorem \ref{NTtheorem} and its corollaries extend to these measures
as well.

\begin{corollary}
      The class of all languages decidable in nondeterministic time at
      most $n(1-\frac{2\lg\lg n}{\lg n})$ infinitely often has 
      $F$-measure 0, $\Gamma_d(\P)$-measure 0, and
      $\Gamma/(\P)$-measure 0.
\end{corollary}
   
   A language $L$ has decision tree depth
   $f(n):\mathbb{N}\longrightarrow\mathbb{N}$ infinitely often if
   $\chi_{L^{=n}}$ has decision tree depth at most $f(n)$ for infinitely many 
   $n$. It is easy to show and well known that a function with decision tree 
   depth $k$ has DNF width at most $k$. See \cite{o2014analysis} for the 
   definition of decision tree depth and a proof of the previous statement. Therefore
   Theorem \ref{NTtheorem} immediately implies the following corollary.
   
   \begin{corollary}
      The set of all languages with decision tree depth at most
      $n(1-\frac{2\lg\lg n}{\lg n})$ infinitely often has $\Gamma(\P)$-measure 
      0.
   \end{corollary}

\bibliographystyle{plain}

\end{document}